\documentclass[twocolumn,showpacs,aps,prd,superscriptaddress]{revtex4}

\usepackage{graphicx}
\usepackage{dcolumn}
\usepackage{amsmath}
\usepackage{epsfig}

\input pubboard/babarsym

\long\def\inst#1{\par\nobreak\kern 4pt\nobreak
    {\it #1}\par\vskip 10pt plus 3pt minus 3pt}
 
\def\dEdx   {\ensuremath{dE/dx}\xspace}
\def\Lc     {\ensuremath{\Lambda_c^+}\xspace}
\def\Lcb    {\ensuremath{\kern  0.18em 
               \overline{\kern -0.18em \Lambda}\kern 0.05em _c^-}\xspace}
\def\pKpi   {\ensuremath{p\Km\pip}\xspace}
\def\LcpKpi {\ensuremath{\Lc \!\!\to\! p\Km\pip}\xspace}
\def\BpKpi  {\ensuremath{\BR_{p\Km\pip}}\xspace}

\def\Xc0    {\ensuremath{\Xi_c^0}\xspace}
\def\pstar  {\ensuremath {p^*}\xspace} 
\def\thstr  {\ensuremath {\theta^*}\xspace} 
\def\cthstr {\ensuremath {\cos\theta^*}\xspace} 
\def\eehad  {\ensuremath{\epem \!\!\to \! {\rm hadrons}}\xspace}
\def\eeqqb  {\ensuremath{\epem \!\!\to \! q\overline{q}}\xspace}
\def\eeccb  {\ensuremath{\epem \!\!\to \! c\overline{c}}\xspace}
\def\NLcqq  {\ensuremath{N_{\Lambda c}^{\qqbar}}\xspace}
\def\NLcyy  {\ensuremath{N_{\Lambda c}^{\Upsilon}}\xspace}
\def\NLcB   {\ensuremath{N_{\Lambda c}^{B}}\xspace}

\newcommand{\BABARPubYear}    {06}
\newcommand{\BABARPubNumber}  {054}

\newcommand{\SLACPubNumber} {12074}
\newcommand{\LANLNumber} {0609004}

\begin{document}

\preprint{\babar-PUB-\BABARPubYear/\BABARPubNumber} 
\preprint{SLAC-PUB-\SLACPubNumber} 

\begin{flushleft}
SLAC-PUB-\SLACPubNumber\\
hep-ex/\LANLNumber\\
\babar-PUB-\BABARPubYear/\BABARPubNumber
\end{flushleft}


\title{\boldmath
Inclusive \Lc Production in $e^+e^-$ Annihilations at $\sqrt{s}=10.54$~\gev \\
and in \Y4S Decays
} 

%
\author{B.~Aubert}
\author{M.~Bona}
\author{D.~Boutigny}
\author{F.~Couderc}
\author{Y.~Karyotakis}
\author{J.~P.~Lees}
\author{V.~Poireau}
\author{V.~Tisserand}
\author{A.~Zghiche}
\affiliation{Laboratoire de Physique des Particules, IN2P3/CNRS et Universit\'e de Savoie,
 F-74941 Annecy-Le-Vieux, France }
\author{E.~Grauges}
\affiliation{Universitat de Barcelona, Facultat de Fisica, Departament ECM, E-08028 Barcelona, Spain }
% ...
\collaboration{The \babar\ Collaboration}
\noaffiliation

\date{\today}%

\begin{abstract}
We present measurements of the total production rates and momentum
distributions of the charmed baryon \Lc 
in \eehad at a center-of-mass energy of 10.54~\gev and in \Y4S decays.
In hadronic events at 10.54~\gev, 
charmed hadrons are almost exclusively leading particles in \eeccb events,
allowing direct studies of $c$-quark fragmentation.
We measure a momentum distribution for \Lc baryons that differs significantly
from those measured previously for charmed mesons.
Comparing with a number of models, 
we find none that can describe the distribution completely.
We measure an average scaled momentum of 
$\left< x_p \right>\! =\! 0.574\pm$0.009
and a total rate of 
$\NLcqq \!\! =\! 0.057\pm$0.002(exp.)$\pm$0.015(BF) \Lc per
hadronic event, where the experimental error is much smaller than that
due to the branching fraction into the reconstructed decay mode, $p\Km\pip$.
In \Y4S decays we measure a total rate of 
$\NLcyy \!\! =\! 0.091\pm$0.006(exp.)$\pm$0.024(BF) per \Y4S decay,
and find a much softer momentum distribution than expected from 
$B$ decays into a \Lc plus an antinucleon and one to three pions.

\end{abstract}


\maketitle

\section{Introduction}

\noindent 
The production properties of charmed baryons in \epem annihilations into \ccbar
and in decays of bottom ($b$) hadrons
probe different aspects of strong interaction physics.
Experiments running at and below the \Y4S resonance are uniquely
positioned to explore each of these processes in detail.
The CLEO experiment has made precise studies of charmed mesons in this 
way~\cite{contdmes}, 
and the larger data samples available at the $B$ factories have allowed 
improved studies of charmed mesons~\cite{bellechsp} and the first 
precise studies of charmed baryons~\cite{bellechsp,bbrxic}.

Heavy hadrons ($H$) produced in \epem annihilations provide a
laboratory for the study of heavy-quark ($Q=c,b$) jet fragmentation, 
in terms of both the relative production rates of hadrons with
different quantum numbers and their associated spectra.
The latter can be characterized in terms of a scaled energy or momentum,
such as $x_{p}$ $\equiv$ $\pstar_H/\pstar_{max}$, 
where $\pstar_H$ is the hadron momentum in the \epem center-of-mass 
(c.m.) frame and $\pstar_{max}\!\! =\!\sqrt{s/4-m_H^2}$ is the 
maximum momentum available to a particle of mass $m_H$ at a c.m.\ energy 
of $\sqrt{s}$.
For $\sqrt{s}\! \gg\! 2m_H$ it has been observed~\cite{pdg} that the $x_p$ 
distributions $P(x_{p})$ for heavy hadrons peak at relatively high values,
and that very few $b$ hadrons are produced apart from those 
containing initial $b$ quarks, so that one can probe leading
$b$-hadron production directly.
This is also the case for charmed ($c$) hadrons when 
$\sqrt{s}<2m_B$, where $m_B$ is the mass of the lightest $b$ meson,
but above \BB threshold a large fraction of the
$c$ hadrons are $b$-hadron decay products.

Since the hadronization process is intrinsically non-perturbative, $P(x_{p})$ 
cannot be calculated using perturbative Quantum Chromodynamics (QCD). 
However, a high quark mass provides a convenient cut-off point and 
the distribution of the scaled momentum of the heavy quark before
hadronization, $x_{Q} \equiv 2\pstar_Q/\sqrt{s}$,
can be calculated~\cite{collins,mn,bcfy,dkt}.
The observable $P(x_{p})$ is thought to be 
related by a simple convolution or hadronization model.
Several phenomenological models of heavy-quark
fragmentation have been proposed~\cite{kart,bowler,pete,lund}. 
Predictions depend on the mass of the heavy quark,
with $P(x_p)$ being much harder for $b$ hadrons than $c$ hadrons, 
and in some cases on the mass and quantum numbers of $H$.
Hadrons containing the same heavy quark type are generally predicted 
to have quite similar $P(x_p)$,
although differences between mesons and baryons have been 
suggested~\cite{kart2,ucla}.
Measurements of $P(x_{p})$ serve to constrain perturbative QCD and 
these model predictions.
Furthermore, measurements for a given $c$ or $b$ hadron at different $\sqrt{s}$
can test QCD evolution, 
and comparisons of $c$- and $b$-hadron distributions can test 
heavy-quark symmetry~\cite{jaffe}. 

The inclusive $b$-hadron scaled energy distribution and its average
value of 0.71 have been measured precisely~\cite{bfrag} by experiments
at the $Z^0$, using partial reconstruction techniques.
However, 
these techniques do not distinguish the different types of $b$ hadrons.
The relative production of $B^-_u$, $B^0_d$, $B^0_s$, 
excited $b$ mesons and
$b$ baryons
have been measured~\cite{lepbrats,pdg}, 
but with limited precision and no sensitivity to differences in their
$x_p$ distributions.
Several $c$ mesons have been studied at the $Z^0$~\cite{lepdmes}, 
but it is difficult to disentangle the leading charm and $b$-decay 
contributions and neither component is measured precisely.
Recent measurements below \BB threshold~\cite{contdmes,bellechsp} 
have good precision over the full $x_p$ range and show substantial
differences between the pseudoscalar $D$ and vector $D^*$ meson states.
$P(x_p)$ has been measured below \BB threshold for two charmed baryons,
\Lc by CLEO~\cite{contcbary} and Belle~\cite{bellechsp}, 
and \Xc0 by \babar~\cite{bbrxic}, 
but with limited statistics, especially at low $x_p$.
In this article we use the excellent particle identification of the
\babar\ experiment to isolate \Lc baryons
(the inclusion of charge conjugate states is implied throughout)
in a large data sample collected below \BB threshold,
at $\sqrt{s}\! =\! 10.54~\gev$.
We measure $P(x_p)$ precisely,
and compare our results with available predictions
and previous measurements of heavy hadrons.

The large $b$-hadron masses allow many hadronic decay modes, 
a small fraction of which have been studied in detail.
The \Y4S\ resonance provides a unique laboratory, in which no $b$~baryons
are produced and decays of mesons into baryons can be studied directly.
Many $c$~baryons have been observed in inclusive \Y4S decays~\cite{pdg}
and the low rate of associated leptons 
and high rate of ``wrong-sign" \Lc ~\cite{bbrbtoccor}
suggest interesting dynamics.
However only a few exclusive decays with $c$ baryons have been
observed~\cite{bellelcp,bellexclc,bellelclck}.
Again, momentum distributions have been measured only for the 
\Lc~\cite{bellechsp,cleoBtocb} and \Xc0 \cite{bbrxic} with limited precision.
Here we use data collected on the \Y4S resonance 
($\sqrt{s}\! =\! 10.58~\gev$) and subtract the \eeccb contribution
measured at $\sqrt{s}\! =\! 10.54~\gev$ to make a precise measurement of the
\Lc momentum distribution in $B$ meson decays,
which we compare with a number of possible models.

In section~\ref{sec:evtsel}, 
we describe the \babar\ detector, in particular the particle
identification capabilities essential to these measurements.
In sections~\ref{sec:lcrec} and~\ref{sec:sextr}, 
we discuss the selection of \Lc candidates and the measurement of
their $x_p$ distributions, respectively. 
We interpret the results for \eeqqb events and \Y4S decays in 
sections~\ref{sec:shape} and~\ref{sec:y4s}, respectively, 
and summarize in section~\ref{sec:summ}.


\section{Apparatus and Event Selection}
\label{sec:evtsel}
 
\noindent
In this analysis, we use data samples corresponding to 9.5~fb$^{-1}$ of 
integrated luminosity at $\sqrt{s}\! =\! 10.54~\gev$
and 81~fb$^{-1}$ on the \Y4S resonance, $\sqrt{s}\! =\! 10.58~\gev$.
The \babar\ detector is located at the PEP-II asymmetric-energy \epem collider
at the Stanford Linear Accelerator Center and is described in detail in 
Ref.~\cite{babarNIM}. 
We use charged tracks measured in the five-layer, double
sided silicon vertex tracker (SVT) and the 40-layer drift chamber (DCH).
In a 1.5~T axial magnetic field, they provide a 
combined resolution on the momentum $p_T$ transverse to the beam axis of
$[\sigma(p_T) / p_T]^2\! =\! [0.0013 p_T]^2 + 0.0045^2$,
where $p_T$ is measured in~\gevc.

Charged particle identification uses a combination of the energy loss (\dEdx)
measured in the DCH, and information from the detector of internally reflected
Cherenkov light (DIRC).
The DCH gas is helium:isobutane 80:20 and a typical cell size is 18 mm.
A truncated mean algorithm gives a \dEdx value for each track with an
average resolution of 7.5\%, from which we calculate a set of five
relative likelihoods $L_i^{\rm DCH}$ for the particle hypotheses 
$i=e, \mu, \pi, K$, and $p$.
Differences between the log-likelihoods 
$l_{ij}^{\rm DCH}= \ln(L_i^{\rm DCH}) - \ln(L_j^{\rm DCH})$
are used as input to the particle identification algorithm.

The DIRC comprises 144 fused silica bars 
that guide Cherenkov photons to an expansion volume filled with water and 
equipped with 10752 photomultiplier tubes.
The fused silica refractive index is 1.473, corresponding to Cherenkov
thresholds of 128, 458 and 867~\mevc for pions, kaons and protons, 
respectively.
A particle well above threshold yields 20--75 measured photons, 
each with a Cherenkov angle resolution of about 10 mrad.
A global likelihood algorithm considers all the reconstructed charged
tracks and detected photons in each event and assigns each track a set 
of likelihoods $L_i^{\rm DIRC}$.

\begin{figure*}[t]
\begin{center} 
\includegraphics[width=\linewidth]{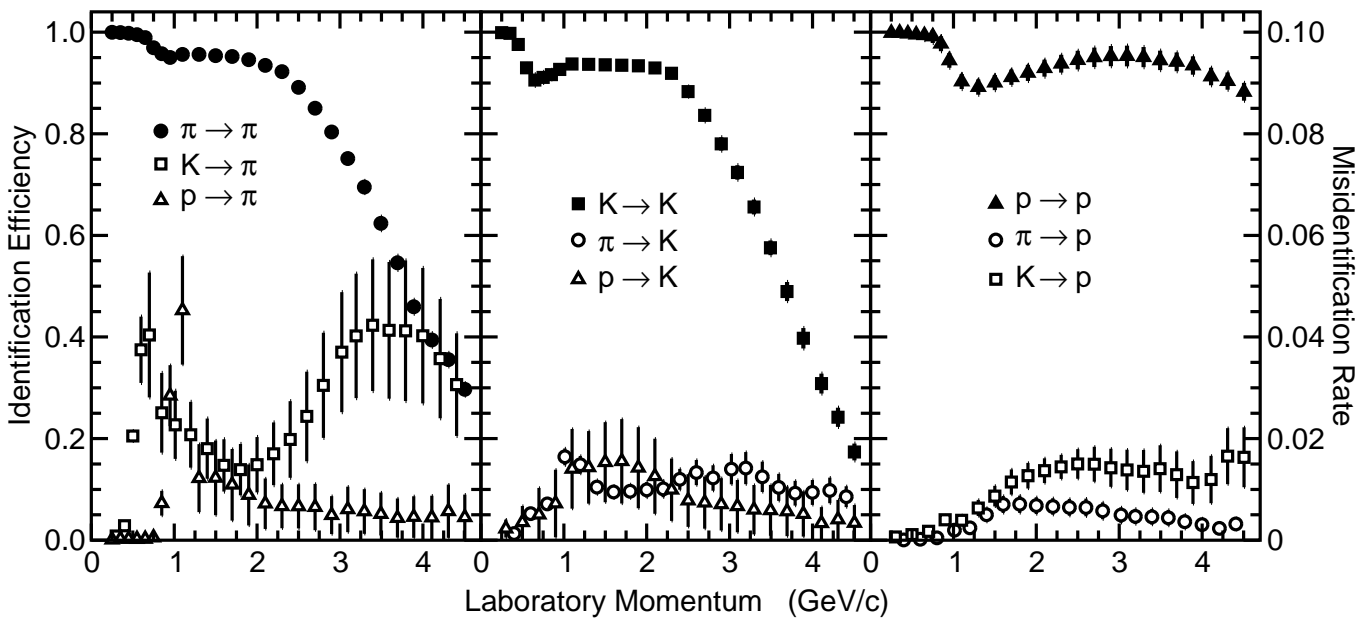}
\end{center}
\vspace{-0.5cm}
\caption {
  Efficiencies for identifying (solid symbols, left-hand vertical scale)
  or misidentifying (open symbols, right-hand scale) selected pions (circles),
  kaons (squares) and protons (triangles) within the DIRC acceptance
  as pions (left), kaons (center) or protons (right).
  They are extracted from control samples in the data as
  smooth functions of momentum and polar angle, and shown in momentum bins
  averaged over the polar angle range used.
  The error bars indicate the average uncertainty due to control
  sample statistics.
  }
\label{fig:pid}
\vspace{-0.29cm}
\end{figure*}

The DCH provides excellent $K$-$\pi$ and $p$-$K$ separation for
momenta in the laboratory frame below 0.5 and 0.9~\gevc, respectively.
The DIRC provides very good separation for momenta above 
1.0 and 1.5~\gevc, respectively.
In both detectors the separation power is lower for tracks with polar angles 
near 90$^\circ$ in the laboratory than for more forward or backward tracks.
To minimize the systematic errors in this analysis, the identification
efficiencies must not vary rapidly as a function of momentum or polar angle.
We have therefore developed an algorithm that uses linear 
combinations of $l_{ij}^{\rm DCH}$ and $l_{ij}^{\rm DIRC}$ chosen to
minimize such variations.
It is described in detail in Ref.~\cite{brandon} and its
performance for tracks used in this analysis is shown as a function of
momentum in Fig.~\ref{fig:pid}.
The identification efficiencies are better than 99\% at low momenta 
and above 90\% for the majority of \Lc decay products.
They are seen to vary smoothly with momentum, and are almost
independent of polar angle except near 0.8~\gevc (1.2~\gevc) for pions
and kaons (protons), where they are as much as 10\% lower for central
tracks than for forward/backward tracks.
The misidentification rates depend strongly on polar angle in the
momentum regions 0.6--0.8, 1.1--1.3 and 2.5--3.5~\gevc.
They are below 5\% everywhere except that the rate for
kaons (protons) to be misidentified as pions reaches 11\%
at 0.7 (1.2)~\gevc for the most central tracks. 
These rates have negligible effects on the results.
About 16\% of the selected tracks have good DCH information but are
outside the DIRC fiducial acceptance.
These can be identified with essentially the same efficiencies as in
Fig.~\ref{fig:pid} for pion and kaon (proton) momenta below 0.6 (0.9)~\gevc.

The event selection requires three or more charged tracks in the event,
which retains any \eeqqb event or \Y4S decay containing a
reconstructable $\Lc \!\!\to\! p\Km\pip$ decay and suppresses beam-related 
backgrounds.
We evaluate its performance using a number of simulations,
each consisting of a generator for a certain type of event 
combined with a detailed simulation of the \babar\ detector~\cite{GEANT}.
For \eeqqb events we use the JETSET~\cite{jetset} generator
and for \Y4S events we use our own generator, EVTGEN~\cite{evtgen},
in which the \Y4S decays into a \BB pair, then the $B$ and \Bb decay
using a combination of measured exclusive and semi-exclusive modes, 
and a $b \!\!\to\! cW^-$ model tuned to the world's inclusive data.
We study large samples of simulated two-photon, $\tau$-pair and 
radiative $e$- and $\mu$-pair events,
and find their contributions to both signal and background
to be negligible.

\section{\boldmath \LcpKpi Selection}
\label{sec:lcrec}

\noindent
We construct \Lc candidates from charged tracks that are consistent with 
originating at the \epem interaction point and have good tracking 
and particle identification information.
Each track must have:
(i) at least 20 measured coordinates in the DCH;
(ii) at least 5 coordinates in the SVT, including at least three in the 
direction along the $e^-$ beam;
(iii) a distance of closest approach to the beam axis below 1~mm;
and (iv) a $z$-coordinate at this point within 10~cm of the nominal
interaction point.
These criteria ensure good quality information from the DCH and a well
measured entrance angle into the DIRC.
If the extrapolated trajectory intersects a DIRC bar then the track is 
accepted if it is identified as a pion, kaon or proton by the combined 
DCH and DIRC algorithm.
If not, then it is accepted if it is identified as a proton (pion or
kaon) using DCH information only and has a momentum below 1.2~(0.6)~\gevc.

\begin{figure}[t]
\begin{center} 
\includegraphics[width=\linewidth]{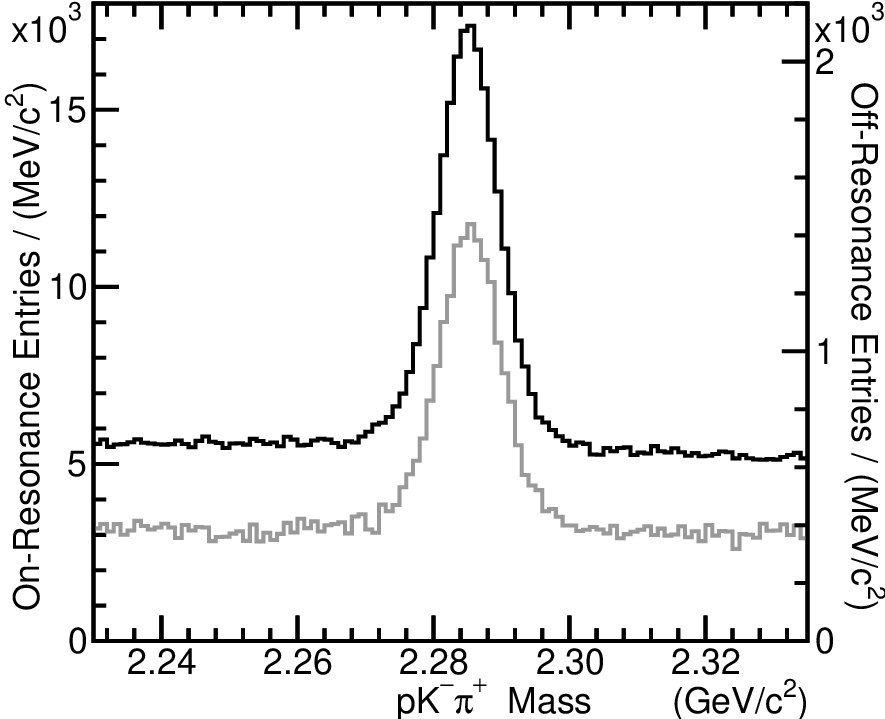}
\end{center}
\vspace{-0.6cm}
\caption {
  Invariant mass distributions for \Lc candidates in the 
  on-(black) and off-resonance (gray) data.
} 
\label{fig:masspkpi}
\vspace{-0.29cm}
\end{figure}

We consider a combination of three charged tracks as a \Lc candidate if
the total charge is $+$1,
one of the positively charged tracks is identified as a proton 
and the other as a pion,
and the negatively charged track is identified as a kaon.  
With the appropriate particle type assigned to each track, we correct their
measured momenta for energy loss and calculate their combined four
momentum from their momenta at their points of closest approach to the
beam axis.
The distributions of invariant mass for the candidates in the on- and
off-resonance data are shown in Fig.~\ref{fig:masspkpi};
\Lc signals of about 137,000 and 13,000 decays, respectively, are
visible over nearly uniform backgrounds.

The \Lc reconstruction efficiency depends primarily on the momenta 
$p$ and polar angles $\theta$ of the daughter tracks in the laboratory
frame.
To reduce systematic uncertainty,
we apply an efficiency correction to each candidate before boosting it
into the \epem c.m.\ frame.
The efficiencies for reconstructing and identifying tracks from pions, kaons 
and protons are determined from large control samples in the
data as two-dimensional functions of ($p$, $\theta$).
We use these efficiencies in dedicated simulations of \qqbar and \Y4S 
events containing a \Lc baryon that is decayed into \pKpi.
From these we calculate the \Lc selection efficiency $\varepsilon$ as
a smooth two-dimensional function of $(p,\theta)$ of the \Lc.
We check that the efficiency does not depend on other track or event
variables, in particular that it is the same in simulated \qqbar and
\Y4S events for given values of $(p,\theta)$.
The resolutions on the \Lc momentum and polar angle are
much smaller than the bin sizes used below, 
so we include resolution effects by defining the efficiency as the
number of \Lc reconstructed within a given $(p,\theta)$ range divided
by the number generated in that range, 
using ranges smaller than the relevant bin sizes.
We test the efficiency using a number of simulations, 
and find biases to be below 1\%.

The efficiency varies rapidly near the edges of the detector
acceptance and at very low momenta in the laboratory.
We make the tight fiducial requirement that
\thstr, the polar angle of the \Lc candidate in the \epem c.m.\ frame, 
satisfy $-$0.7$<\! \cthstr\!\! <$0.2,
which reduces model dependence and rejects all candidates in regions 
with efficiency below about 5\%, 
including those with a total laboratory momentum below about 0.7~\gevc.
A feature of the boosted c.m.\ system is that true \Lc baryons with low
c.m.\ momentum \pstar are boosted forward and often have all three
tracks in the detector acceptance, 
giving efficient access to the full \pstar range.

We define $A_k$ as the fraction of the \Lc in events of type $k$ 
produced within our fiducial range $-$0.7$<\! \cthstr\!\! <$0.2.
In $\Y4S \!\to\! \BB \!\to\! \Lc X$ decays, 
the true \cthstr distribution is uniform and $A_\Upsilon\! =\! 0.45$.
In \ccbar events the angular distribution of the initial $c$-quark follows 
$1+\cos^2\theta^*$, and we use the JETSET simulation to calculate the 
distribution for \Lc after QCD radiation and hadronization.
Soft \Lc are produced predominantly in events with hard gluon
radiation, which flattens the distribution considerably.
The resulting value of $A_{\ccbar}$ is 0.46 at $\pstar\!\! =\! 0$, 
and falls with increasing \pstar toward an asymptotic value of 0.38.

\begin{figure}[t]
\begin{center} 
 \includegraphics[width=\linewidth]{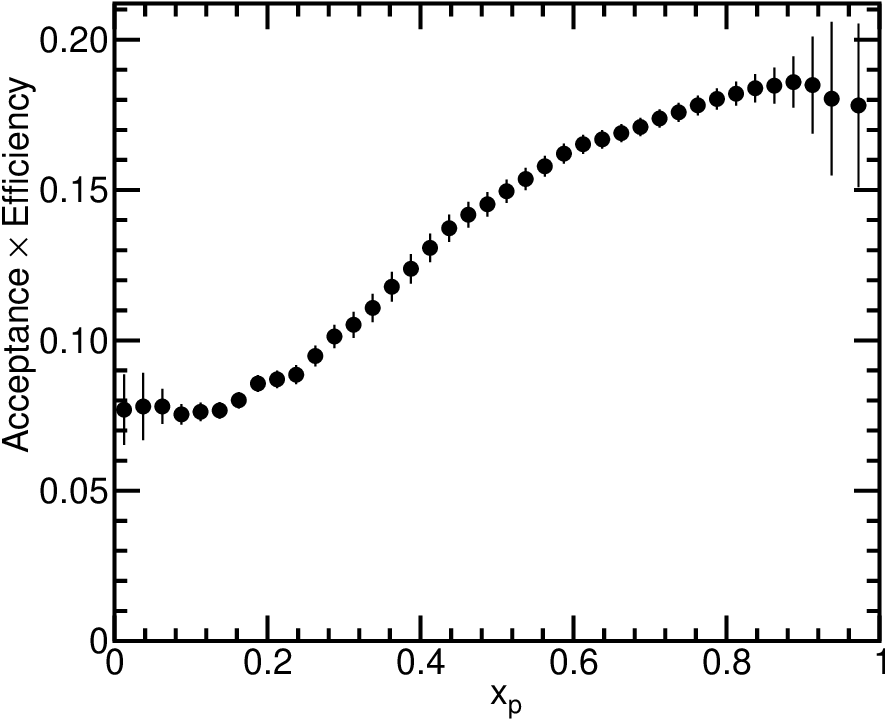}
\end{center}
\vspace{-0.6cm}
\caption {
  Average value of $A_{\ccbar}\cdot\varepsilon$, the 
  $\Lc \!\!\to\! p\Km\pip$ acceptance times reconstruction efficiency,
  for candidates in the off-resonance data in bins of scaled momentum $x_p$.
  The error bars represent the average statistical uncertainty.
  }
\label{fig:effpkpi}
\end{figure}

We bin candidates according to their reconstructed values of 
$x_p\! = \pstar/\pstar_{max}$, where 
$\pstar_{max}$ is calculated for each event from its c.m.\ energy 
and the nominal \Lc mass~\cite{pdg}.
Figure~\ref{fig:effpkpi} shows the average value in each $x_p$ bin of
the product $A_{\ccbar}(\pstar)\cdot \varepsilon(p,\theta)$ for 
selected candidates in the off-resonance data.
It ranges from 8\% at low $x_p$ to 19\% at high $x_p$.
The error bars represent the statistical uncertainty on
the efficiency calculation.
The corresponding quantity for \Lc from \Y4S decays, 
$\left< A_\Upsilon\cdot\varepsilon(p,\theta)\right>$, 
is slightly higher at low $x_p$ due to a 
small dependence of $\varepsilon$ on $\theta$,
and rises faster with increasing $x_p$ since $A_\Upsilon$ is constant,
whereas $A_{\ccbar}$ decreases.
We give each candidate a weight equal to the inverse of either 
$A_{\ccbar}\cdot\varepsilon$ or $A_\Upsilon\cdot\varepsilon$,
as specified below.
The RMS deviation of the weights in each bin is always much smaller than the 
average value.

\section{Signal Extraction}
\label{sec:sextr}

\noindent
To estimate the number of $\Lc \!\!\to\! p\Km\pip$ decays in each $x_p$ bin 
in the data,
we fit the weighted invariant mass distribution 
with a function comprising signal and background components.
Based on the simulated mass distributions, 
we describe the signal with a sum of two Gaussian functions of
common mean value, one of which has 1.5 times the width and one quarter
of the area of the other,
and correct for a 1.3\% residual bias in the fitted area.
We check the simulated bias by comparing with a single Gaussian signal 
function.
The change in the yields is 1.2\% in both data and simulation.

The simulation predicts a nearly uniform background over the \pKpi
mass range shown in Fig.~\ref{fig:masspkpi}.
We search for reflections in the data by changing the particle mass 
assignments.
We observe signals for 
$D^+   \!\!\to\! \Km\pip\pip$ (\pip misidentified as $p$) and 
$D_s^+ \!\!\to\! \Km\Kp\pip$ ($K^+$ misidentified as $p$)
at very low levels consistent with the predictions of our detector
simulation, but no unexpected structure.
From these studies we calculate that reflections known to give broad
structures in the vicinity of the \Lc peak, such as
$D^{*+} \!\!\to\! \Km\pip\pip$ (\pip misidentified as $p$)
contribute a number of entries in each bin much smaller than the
statistical fluctuations.
We also study processes such as
$\Sigma_c^{++} \!\!\to\! \Lc\pip$, with the wrong \pip included in the
\Lc candidate, and find their contributions to be negligible.
In each $x_p$ bin, a linear function describes the mass distribution
in the data over a wide range away from the \Lc peak region,
so we use a linear background function and perform fits over the
range 2235--2335~\mevcc.

\begin{figure}[t]
\begin{center} 
 \includegraphics[width=\linewidth]{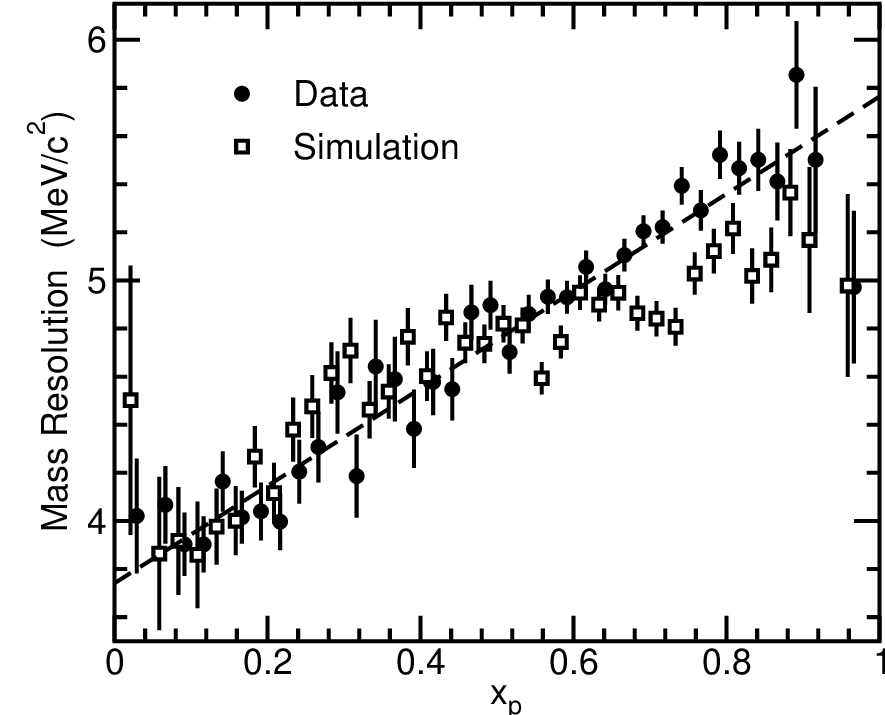}
\end{center}
\vspace{-0.6cm}
\caption {
  RMS width of the fitted \Lc signal function in the data (circles) 
  and simulation (squares) as a function of $x_p$.
  The line represents a linear parametrization of the data.
  }
\label{fig:mwpkpi}
\end{figure}

We first fit the full data sample in each $x_p$ bin in order to study 
mass resolution and bias.
These fits yield \Lc mass values that vary slightly with $x_p$ in a
manner consistent with the simulation and 
our recent measurement of the \Lc mass~\cite{lcmass}.
The fitted mass resolutions 
(RMS width of the signal function) 
are shown as a function of $x_p$ in Fig.~\ref{fig:mwpkpi}.
The simulation is consistent with the data at low $x_p$, and is
slightly optimistic at high $x_p$.
The effect of this difference on the efficiency estimate is negligible.

\begin{figure}[t]
\begin{center} 
 \includegraphics[width=\linewidth]{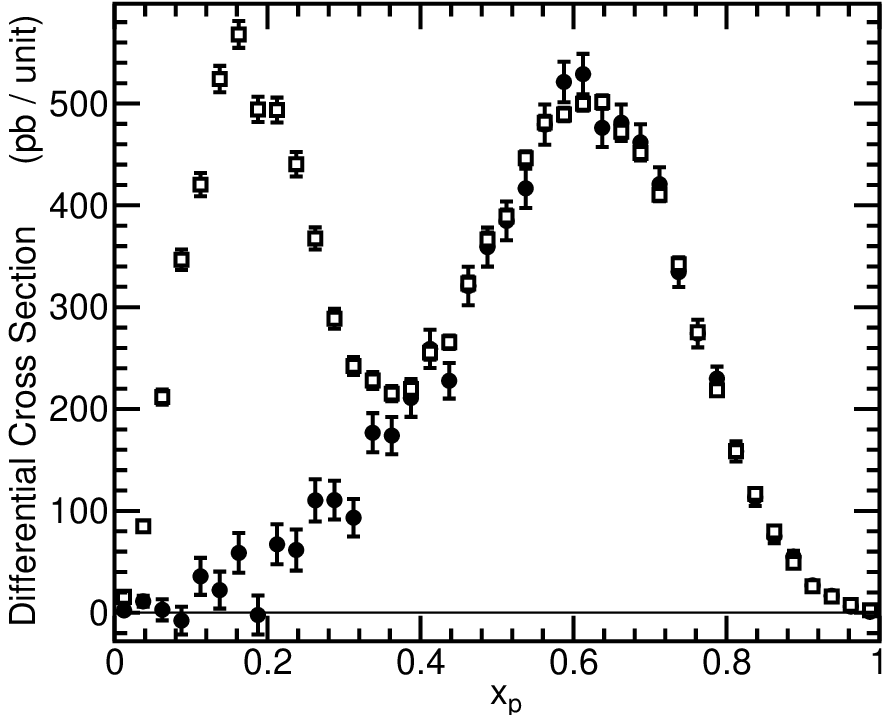}
\end{center}
\vspace{-0.6cm}
\caption {
  Differential cross sections for 
  $\Lc\! + \Lcb$ production
  in the off-(circles) and on-resonance (open squares) data 
  as functions of $x_p$.
  The errors are statistical only. 
  } 
\label{fig:xsonofpkpi}
\end{figure}

Next we fix the mean and width of the signal function in each $x_p$ bin
to values from linear parametrizations,
and perform fits to the on- and off-resonance data separately.
Dividing the signal yields by the integrated luminosity, bin width and 
branching fraction 
$\BpKpi \!\equiv\! \BR(\LcpKpi )\! =\! 5.0\pm1.3$\%~\cite{pdg}
gives the differential production cross sections
shown in Fig.~\ref{fig:xsonofpkpi} 
with statistical errors only.
We use the \eeccb acceptance factor $A_{\ccbar}$ for the off-resonance data
and, for purposes of this comparison, an average value of $A$ in each
bin weighted by the relative production rates measured below 
(see Tables~\ref{tab:xscont} and~\ref{tab:xsy4s})
for the on-resonance data.
There are two broad peaks in the on-resonance cross section, 
corresponding to the contributions from \Y4S decays at low $x_p$ and from 
\eeccb events at high $x_p$.
For $x_p\! >\! 0.47$, 
the kinematic limit for a $B$ decay including a \Lc and an antiproton,
the two cross sections are consistent, indicating no visible contribution from 
\Y4S events.

We extract the cross section for \Y4S decays by repeating the
analysis using $A_\Upsilon$ for both data sets.
In each $x_p$ bin we then subtract the off-resonance cross section,
scaled down by 0.8\% to account for the dependence of the cross
section on the c.m.\ energy,
from the on-resonance cross section.
We divide the off-resonance and \Y4S cross sections by the  
\eehad and effective $e^+e^- \!\!\to\! \Y4S$ cross sections, respectively, 
to yield the differential production rates per event 
discussed in the following sections.

We consider several
sources of systematic uncertainty.
We propagate the uncertainties on the measured track finding and particle
identification efficiencies by recalculating
$\varepsilon(p,\theta)$ with all efficiencies of a given type varied
simultaneously and repeating the fits.
Tracking gives an uncertainty of 2.5\% and the particle identification
contributions total 1.5--2.1\%, depending on $x_p$.
We obtain a 0.9\% uncertainty due to the resonant substructure of the \LcpKpi
decay similarly.
The uncertainty on our integrated luminosity is 1.0\%.
Simulation statistics contribute 2--4\% where the rate is
significantly nonzero.
We check the fitting procedure by
floating the signal mean and/or width,
fixing them to nominal or fitted values,
using a single Gaussian signal function, 
using a quadratic background function, 
and varying the bin size.
All changes in the signal yields are less than the
corresponding statistical errors,
and we take the largest change in each $x_p$ bin as a
systematic uncertainty.
Each simulated bias is varied by $\pm$50\%; together they contribute
0.7\% to the systematic uncertainty.
The imperfect $x_p$ distribution used in the efficiency calculation
can affect the result if the efficiency varies over the width of an
$x_p$ bin.
We recalculate the efficiency with the input distribution shifted by
plus and minus our bin width to derive a conservative limit
on any such effect of 0.5--1.9\%, depending on $x_p$, which we take as a
systematic uncertainty.

We also perform several systematic checks of the results.
The cross sections measured separately for \Lc and \Lcb,
which have very different efficiencies at low laboratory momentum, 
are consistent.
Cross sections measured in six different regions of \cthstr are
consistent with each other.
Due to the boosted c.m.\ system, these would be affected differently by any
deficiency in the detector simulation, especially at low $x_p$.
They also have very different $A_{\ccbar}$ values, 
and these studies indicate that the uncertainties due to both
the production angle model in \ccbar events and the value of the boost
are negligible.

\section{\boldmath Results for $\sqrt{s}=10.54$ \gev}
\label{sec:shape}

\subsection{\boldmath \Lc Baryon Production}

\begin{figure}[t]
\begin{center} 
\includegraphics[width=\linewidth]{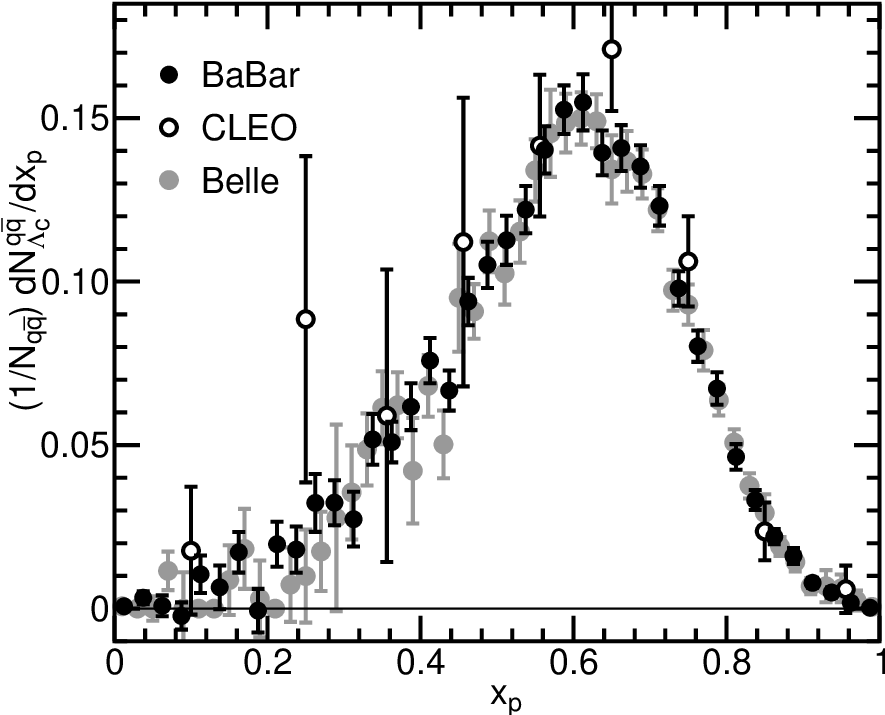}
\end{center}
\vspace{-0.6cm}
\caption {
  Differential \Lc production rate per \eeqqb event 
  compared with previous measurements.
  The error bars include statistics and those systematic errors that
  affect the shape.
  Each experiment has a normalization uncertainty of a few percent,
  and there is an overall 26\% uncertainty due to the
  \LcpKpi branching fraction.
  } 
\label{fig:xscont}
\end{figure}

\noindent
The differential \Lc production rate per hadronic event
$(1/N_{\qqbar})(d\NLcqq / dx_p)$ is
tabulated in Table~\ref{tab:xscont} and compared with previous
charmed baryon measurements in Fig.~\ref{fig:xscont}.
We distinguish between systematic uncertainties that affect the shape 
of the cross section and those that affect only its normalization.
The former include both uncertainties that are uncorrelated between bins
and the parts of the correlated uncertainties whose values depend on $x_p$.
The uncertainty from the fitting procedure has negligible correlation
between bins,
and those from the particle identification and the shift of the
simulated distribution have only very short-range correlations,
so we include them in the uncorrelated category.
We express the error matrix for the efficiency calculation as the sum of
a diagonal ``uncorrelated" matrix and a remainder matrix, in which the
elements of the former are as large as possible but no correlation
coefficient in the latter exceeds unity.
The sum in quadrature of the uncorrelated uncertainties is listed in 
the ``independent" column of Table~\ref{tab:xscont}.
It is typically 3\% in the peak region, increasing to 10\% 
where the cross section is one-third of its peak value, and
becoming relatively large at the ends of the $x_p$ range.
The square roots of the diagonal elements of the remainder matrix 
are listed in the ``correlated" column of Table~\ref{tab:xscont},
and included with the independent and statistical components in the
error bars in the figures.

\begin{table}[htb]
\caption{ \label{tab:xscont} 
  \Lc differential production rate per hadronic event per unit
  $x_p$ at $\sqrt{s}\! =\! 10.54$~\gev.
  The last column includes experimental errors that are
  correlated between $x_p$ values and affect the shape of the distribution.
  Normalization uncertainties are given only on the total.
  }
\begin{center}
\begin{tabular}{cr@{\hspace{0.4cm}}ccc}
\hline\hline
       &      &             &              &                \\[-0.3cm]
 $x_p$ &      & Statistical & \multicolumn{2}{c}{Systematic}\\[-0.35cm]
 Range & \multicolumn{1}{c}{\Large $\frac{1}{N_{\qqbar}}
                            \frac{d\NLcqq}{dx_p}$} &
 Error & \hspace*{-0.1cm}Independent         & Correlated \\[0.25cm]
\hline
              &          &          &          &          \\[-0.3cm]
 0.000--0.025 &   0.0007 &   0.0009 &   0.0004 &   0.0001 \\
 0.025--0.050 &   0.0033 &   0.0016 &   0.0012 &   0.0002 \\
 0.050--0.075 &   0.0008 &   0.0030 &   0.0011 &   0.0001 \\
 0.075--0.100 &$-$0.0023 &   0.0040 &   0.0013 &   0.0001 \\
 0.100--0.125 &   0.0105 &   0.0053 &   0.0020 &   0.0007 \\
 0.125--0.150 &   0.0065 &   0.0053 &   0.0041 &   0.0002 \\
 0.150--0.175 &   0.0172 &   0.0057 &   0.0024 &   0.0004 \\
 0.175--0.200 &$-$0.0006 &   0.0056 &   0.0036 &   0.0000 \\
 0.200--0.225 &   0.0197 &   0.0057 &   0.0038 &   0.0004 \\
 0.225--0.250 &   0.0180 &   0.0059 &   0.0039 &   0.0003 \\
 0.250--0.275 &   0.0323 &   0.0061 &   0.0064 &   0.0006 \\
 0.275--0.300 &   0.0324 &   0.0056 &   0.0040 &   0.0006 \\
 0.300--0.325 &   0.0273 &   0.0054 &   0.0064 &   0.0006 \\
 0.325--0.350 &   0.0517 &   0.0056 &   0.0053 &   0.0011 \\
 0.350--0.375 &   0.0509 &   0.0053 &   0.0029 &   0.0010 \\
 0.375--0.400 &   0.0617 &   0.0054 &   0.0045 &   0.0012 \\
 0.400--0.425 &   0.0759 &   0.0055 &   0.0040 &   0.0014 \\
 0.425--0.450 &   0.0667 &   0.0051 &   0.0032 &   0.0010 \\
 0.450--0.475 &   0.0939 &   0.0055 &   0.0044 &   0.0014 \\
 0.475--0.500 &   0.1051 &   0.0056 &   0.0041 &   0.0014 \\
 0.500--0.525 &   0.1126 &   0.0056 &   0.0048 &   0.0014 \\
 0.525--0.550 &   0.1220 &   0.0056 &   0.0043 &   0.0015 \\
 0.550--0.575 &   0.1403 &   0.0058 &   0.0041 &   0.0014 \\
 0.575--0.600 &   0.1526 &   0.0058 &   0.0044 &   0.0015 \\
 0.600--0.625 &   0.1548 &   0.0058 &   0.0061 &   0.0014 \\
 0.625--0.650 &   0.1394 &   0.0055 &   0.0038 &   0.0012 \\
 0.650--0.675 &   0.1409 &   0.0052 &   0.0045 &   0.0012 \\
 0.675--0.700 &   0.1352 &   0.0052 &   0.0037 &   0.0011 \\
 0.700--0.725 &   0.1232 &   0.0049 &   0.0035 &   0.0010 \\
 0.725--0.750 &   0.0979 &   0.0043 &   0.0030 &   0.0009 \\
 0.750--0.775 &   0.0803 &   0.0040 &   0.0026 &   0.0007 \\
 0.775--0.800 &   0.0673 &   0.0034 &   0.0035 &   0.0008 \\
 0.800--0.825 &   0.0464 &   0.0029 &   0.0026 &   0.0006 \\
 0.825--0.850 &   0.0332 &   0.0025 &   0.0017 &   0.0004 \\
 0.850--0.875 &   0.0220 &   0.0020 &   0.0010 &   0.0005 \\
 0.875--0.900 &   0.0161 &   0.0017 &   0.0017 &   0.0006 \\
 0.900--0.925 &   0.0079 &   0.0013 &   0.0007 &   0.0003 \\
 0.925--0.950 &   0.0049 &   0.0012 &   0.0006 &   0.0008 \\
 0.950--0.975 &   0.0018 &   0.0006 &   0.0003 &   0.0004 \\
 0.975--1.000 &   0.0003 &   0.0001 &   0.0002 &   0.0001 \\
\hline
              &          &          &          &          \\[-0.3cm]
 Total        &   0.0568 &   0.0007 &   0.0006 &   0.0008 \\
 Norm. err.   &   0.0016 &          &          &          \\
 BF error     &   0.0148 &          &          &          \\
\hline\hline
\end{tabular}
\end{center}
\end{table}

All other uncertainties are fully correlated between bins and very
nearly independent of $x_p$, so are considered experimental normalization 
uncertainties.
They total 2.9\%, dominated by the track-finding efficiency.
There is a 26\% uncertainty on \BpKpi that also affects the
normalization.
The integral of the differential rate, taking the correlation in the errors 
into account, gives the total rate, 
listed at the bottom of Table~\ref{tab:xscont} 
along with the normalization uncertainties.
The product of the total rate per event and branching fraction of 
$\NLcqq \cdot \BpKpi = 2.84 \pm 0.04\,\, {\rm (stat.)}
                            \pm 0.09\,\, {\rm (syst.)} \times 10^{-3}$
is consistent with, and more precise than, previous measurements.
The normalization uncertainties are not included in any of the figures,
and all rates shown assume the same value of \BpKpi.

Assuming the \Lc are produced predominantly in \eeccb events, the
total rate corresponds to a rate of
$N_{\Lambda c}^c\!\! = \! 0.071\pm$0.003 (exp.)$\pm$0.018 (BF) \Lc 
per $c$-quark jet.
Roughly 10\% of the particles in high-energy jets have generally been
observed to be baryons~\cite{pdg}, 
and our measurement is consistent with 10\% of $c$~jets producing a
$c$~baryon, with a large fraction of these decaying via a \Lc.
All known non-strange charmed baryons decay predominantly through a \Lc, 
whereas no $\Omega_c$ states and only the heaviest observed $\Xi_c$
states~\cite{bellexic} are known to decay through a \Lc, 
so that 70--85\% of inclusive charmed baryons would be expected to 
decay through a \Lc in current hadronization models.

\begin{figure}[t]
\begin{center} 
\includegraphics[width=\linewidth]{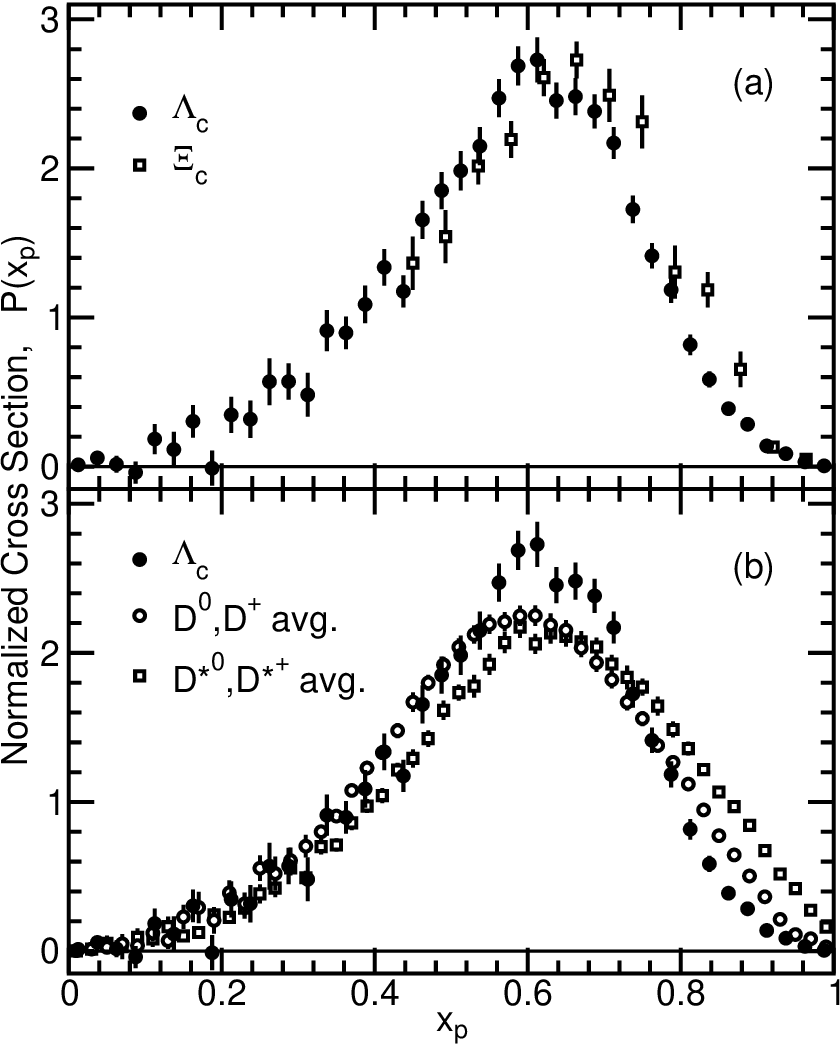}
\end{center}
\vspace{-0.6cm}
\caption {
  $x_p$ distribution for \Lc in \eeqqb events compared with those measured 
  for (a) \Xc0 by \babar~\cite{bbrxic} and
  (b) charmed mesons by Belle~\cite{bellechsp}.
  }
\label{fig:xsdlc}
\end{figure}

The shape of the differential production rate is consistent with
previous results and measured more precisely.
It is quite hard, as expected, peaking near $x_p\! = \! 0.6$.
We normalize the rate to unit area to obtain $P(x_p)$,
and compare it with previously measured distributions for 
\Xc0 baryons and $D$ and $D^*$ mesons in Fig.~\ref{fig:xsdlc}.
The \Xc0 distribution is normalized to have the same peak height 
as the \Lc distribution,
and, since it was measured on the \Y4S resonance, 
is shown only above the kinematic limit for $B$-meson decays.
The two charmed baryons have similar distributions,
with that for the heavier baryon shifted up in $x_p$ by roughly 0.05.
Although qualitatively similar, the $D$ meson distributions show
broader peaks than the baryon distributions and differ greatly in the 
way they fall toward zero at high $x_p$.
The charmed baryon and meson distributions are all much softer than
the inclusive $B$-hadron distribution at c.m.\ energies well above
\bbbar threshold, which peaks around $x_p\! =\! 0.75$~\cite{bfrag}.

The average $x_p$ value is often used in comparisons between different 
heavy hadrons, and the higher moments of the distribution are of
theoretical interest.
In Table~\ref{tab:xsmomts} we list values of the first six moments of
the $x_p$ distribution, calculated by summing over bins.
They are consistent with previous measurements from Belle~\cite{bellechsp};
all are 1--2 standard deviations lower,
but the moments are strongly correlated with each other.
The $\left< x_p \right>$ value of 0.574$\pm$0.009 is consistent with
those measured~\cite{bellechsp} for $D^0$ and $D^+$ mesons, and about 5\%
lower than those for $D^{*0}$ and $D^{*+}$ mesons.

\begin{table}[t]
\caption{ \label{tab:xsmomts}
  The first six moments of the \Lc $x_p$ distribution in hadronic
  events at $\sqrt{s}\! =\! 10.54$ \gev.
  }
\begin{center}
\begin{tabular}{cc@{\hspace*{0.3cm}}c@{\hspace*{0.3cm}}c
                  @{\hspace*{0.3cm}}c@{\hspace*{0.5cm}}c}
\hline\hline
        &      &          &        &        &                   \\[-0.3cm]
        &      & Stat.& \multicolumn{2}{c}{Systematic\hspace*{0.6cm}}  &   \\
 Moment & Value& Error    & Indep. & Correl.& Belle             \\[0.05cm]
\hline
    &          &          &        &        &                   \\[-0.3cm]
$\left< x_p \right>$ 
    &  0.5738  &  0.0061  & 0.0049 & 0.0032 & 0.5824$\pm$0.0025 \\[0.15cm] 
$\left< x_p^2 \right>$ 
    &  0.3544  &  0.0038  & 0.0030 & 0.0021 & 0.3649$\pm$0.0034 \\[0.15cm] 
$\left< x_p^3 \right>$ 
    &  0.2305  &  0.0026  & 0.0021 & 0.0015 & 0.2396$\pm$0.0023 \\[0.15cm] 
$\left< x_p^4 \right>$ 
    &  0.1560  &  0.0020  & 0.0015 & 0.0011 & 0.1630$\pm$0.0051 \\[0.15cm] 
$\left< x_p^5 \right>$ 
    &  0.1090  &  0.0015  & 0.0012 & 0.0009 & 0.1151$\pm$0.0020 \\[0.15cm] 
$\left< x_p^6 \right>$ 
    &  0.0783  &  0.0012  & 0.0010 & 0.0008 & 0.0851$\pm$0.0023 \\[0.15cm] 
\hline
\end{tabular}
\end{center}
\end{table}

\subsection{\boldmath Tests of $c$-Quark Fragmentation Models}
\label{subsec:model}

\noindent
Testing models of heavy-quark fragmentation can be problematic since 
the predictions are usually functions of a variable $z$ that is
not accessible experimentally,
such as
$z_1=(E+p_{\|})_{H}/(E+p_{\|})_Q$, $z_2 = {p_{\|}}_H / {p_{\|}}_Q$ or
$z_3=p_{H} /p_{Hmax}(p_{Q})$,
where $p_{\|}$ represents a momentum projection on the 
flight direction of the heavy quark before it hadronizes.
Monte Carlo event generators use similar internal variables, 
and in some cases can be made to produce events according to a given
input function $f(z,\beta)$, 
where $\beta$ represents the set of model parameters.
In this way one can test the large-scale features of any model,
although the detailed structure may not be reproduced exactly.

We consider the perturbative QCD calculations of 
Collins \& Spiller (CS)~\cite{collins} and
Braaten et al.~(BCFY)~\cite{bcfy},
as well as the phenomenological models of 
Kartvelishvili et al.\ for mesons (KLP-M)~\cite{kart} 
and baryons (KLP-B)~\cite{kart2},
Bowler~\cite{bowler},
Peterson et al.~\cite{pete}, 
the Lund group~\cite{lund}, 
the UCLA group~\cite{ucla} and the HERWIG group~\cite{herwig}.
The latter two include heavy quark fragmentation within their own
generators, and
the other seven predict the functional forms listed in 
Table~\ref{tab:fragmodels}.
We implement each of these functions $f(z,\beta)$ within 
the JETSET generator.
JETSET uses $z_1$ as its internal variable, but $z_2$ and $z_3$
are very similar at high $x_p$ where we are most sensitive to the
shape.
All distributions are affected by JETSET's simulation of
hard and soft gluon radiation.

\begin{table}[t]
\caption{ \label{tab:fragmodels} 
  Fragmentation models compared with the data.
  Here $m_{\perp}^2\! =\! m_H^2 + p_\perp^2$, $p_\perp$ is the
  component of the hadron momentum transverse to the quark momentum,
  and the $z_i$ are defined in the text.
  }
\begin{center}
\begin{tabular}{lr@{}lc}
\hline\hline
         &                 &                         &               \\[-0.3cm]
Model    &  \multicolumn{2}{c}{$f(z,\beta)$}         &  Ref.         \\[0.05cm]
\hline
         &                 &                         &               \\[-0.2cm]
CS       & $([1-z_1]/z_1 + \epsilon$
         & $[2-z_1]/[1-z_1])\;(1+z_1^2)\;\times$     & \cite{collins}\\
        && $(1- 1/z_1 - \epsilon/[1-z_1])^{-2}$      &               \\[0.15cm]
BCFY     & $z_2 (1-z_2)^2$     & $(1-[1-d]z_2)^6\;\times$&\cite{bcfy}\\
         & $[\,3 -3z_2  \cdot$ & $(\;\;3-\;\;4d)$    &               \\
         & $+\,\;\;z_2^2\cdot$ & $(   12-23d+26d^2)$ &               \\
         & $-\,\;\;z_2^3\cdot$ & $(\;\;9-11d+12d^2)(1-d)$ &          \\
         & $+\,   3z_2^4\cdot$ & $(\;\;1-\;\;\;\,d+
                                         \;\;\;\,d^2)(1-d)^2 \,]$  & \\[0.15cm]
KLP-M    & $z_2^\alpha \cdot$  & $(1-z_2)$           & \cite{kart}   \\[0.15cm]
KLP-B    & $z_2^\alpha \cdot$  & $(1-z_2)^3$         & \cite{kart2}  \\[0.15cm]
Bowler   & $z_3^{-(1+bm_{\perp}^{2})}$
         & $(1-z_3)^{a}\exp(-bm_{\perp}^{2}/z_3)$    & \cite{bowler} \\[0.15cm]
Peterson & $(1/z_2)$
         & $(1-[1/z_2]- \epsilon/[1-z_2])^{-2}$      & \cite{pete}   \\[0.15cm]
Lund     & $(1/z_1)$  
         & $(1-z_1)^{a}\exp(-bm_{\perp}^{2}/z_1)$    & \cite{lund}   \\[0.1cm]
\hline\hline
\end{tabular}
\end{center}
\end{table}

We test each model against our measured $P(x_p)$ using a binned $\chi^2$
\begin{equation}
\chisq = \sum_{i,j=1}^{n} (P_{i}^{data} - P_{i}^{MC}) {\bf V}^{-1}_{i,j}
                          (P_{j}^{data} - P_{j}^{MC}) 
\label{eqn:chisq}
\end{equation}
where $n$ is the number of bins, 
$P_{i}^{data}$ ($P_{i}^{MC}$) is the fraction of the measured (modelled) 
distribution in bin $i$, and
${\bf V}$ is the full error matrix formed from the errors on the data
(Table~\ref{tab:xscont}) with the statistical errors on the simulation
added in quadrature to the diagonal elements.
For each function in Table~\ref{tab:fragmodels} we minimize this 
\chisq with respect to the set of parameters $\beta$ by
scanning over a wide range of possible $\beta$ values and generating a
large sample of \qqbar events at each of several points.
We then generate additional sets near each minimum.
All other model parameters are fixed to their default values.

The functions with one free parameter (CS, BCFY, KLP-M, KLP-B and Peterson) 
all show a single, well behaved minimum.
The two parameters in the Bowler and Lund functions are strongly
correlated, and the \chisq shows a single narrow valley.
The UCLA model has internal parameters $a$ and $b$ that control production 
of all particles simultaneously; 
since it has been suggested that their values should be different for 
mesons and baryons, we vary them by the same procedure, again finding
a strong correlation and a narrow \chisq valley.
HERWIG has no free parameter controlling heavy hadron production, so
we consider only the default parameter values.

\begin{figure*}[t] 
\begin{center} 
\includegraphics[width=0.98\linewidth]{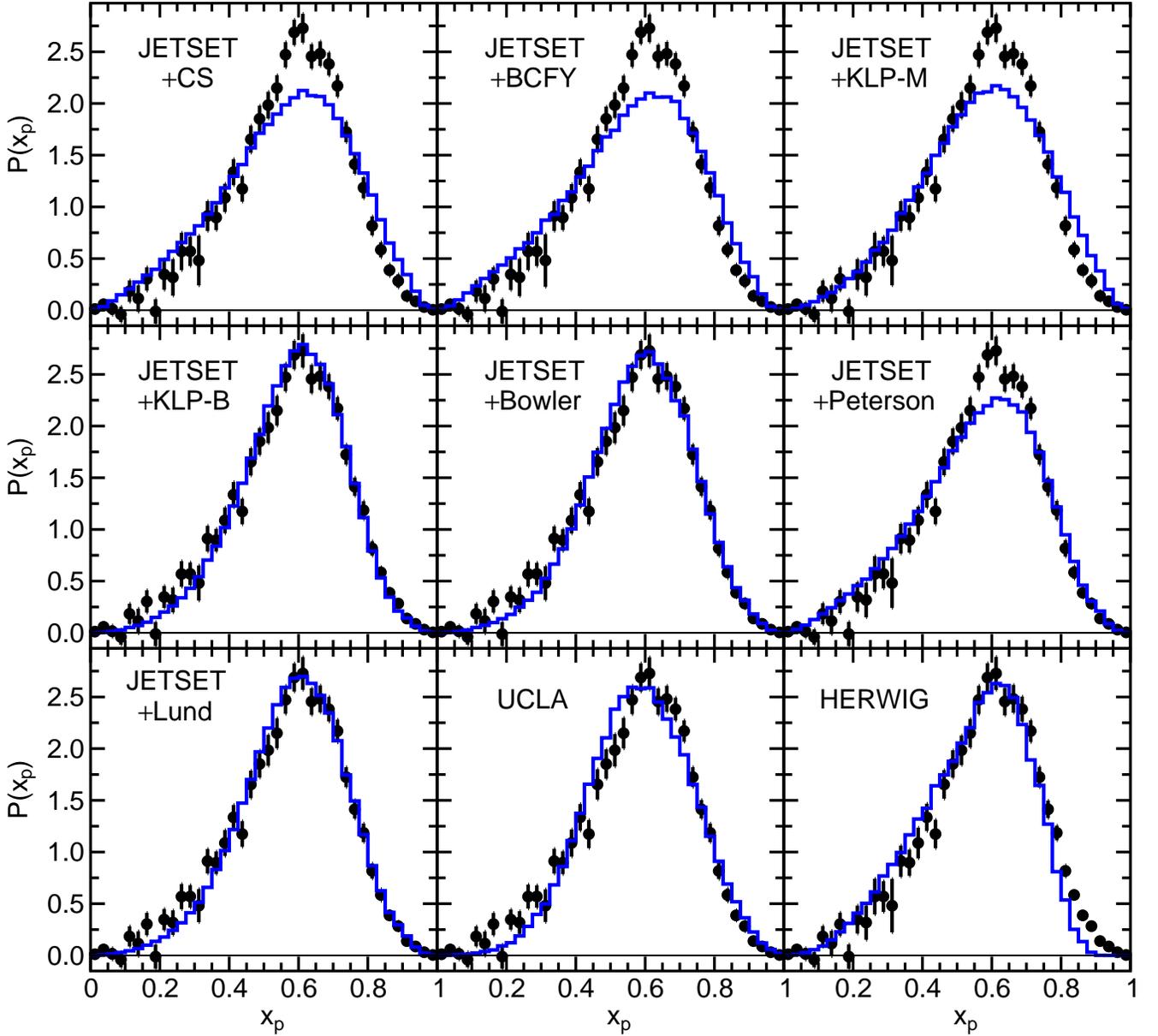}
\end{center} 
\vspace{-0.7cm}
\caption {
  The \Lc $x_p$ distribution (dots) compared with the results of the 
  model tests (histograms) described in the text.
  The error bars include simulation statistics.
  }
\label{fig:fragmodel}
\end{figure*}   

We compare the fitted distributions with the data in 
Fig.~\ref{fig:fragmodel}, and
list the $\chi^2$, the fitted parameter values and the average $x_p$
of each fitted distribution in Table~\ref{tab:modelresult}.
The parameter values are meaningful only in the context of the JETSET
(or UCLA) model.
The KLP-B, Lund and Bowler models give the best descriptions of the data, 
with respective \chisq confidence levels of 0.15, 0.11 and 0.06.
However their fitted distributions are systematically below the data at the 
lowest $x_p$ values and above the data just below the peak region.
The UCLA distribution is qualitatively similar to these three models
but falls more rapidly at low $x_p$, resulting in poor agreement with
the data.
The CS, BCFY and KLP-M models predict distributions that are much too broad,
and the Peterson distribution is also too broad.
The HERWIG distribution is consistent with the data in the peak region, 
but cuts off too sharply at high $x_p$.

The fitted values of the parameter $a$ for the UCLA, JETSET$+$Lund and
JETSET$+$Bowler models are larger than those that describe the
production of inclusive light hadrons and charmed mesons
($a\!\approx\!1.2$ for UCLA and $0.1\!\leq\! a \!\leq\! 0.6$ 
for the other two models).
Differences between baryon and meson distributions have been suggested
on the basis of quark counting~\cite{kart2,ucla}.
Crossing of the diagram for leading hadron production in \epem
annihilation gives a deep inelastic scattering diagram, 
calculations for which depend on the number of spectator quarks, $N_s$;
in the limit $z\! \rightarrow$1, one expects 
$f(z)\! \propto\! (1-z)^{2N_s-1}$~\cite{DIS,kart2}, 
and $2N_s-1\! =\! 3$ for baryons.
This is the form of the KLP-B function, 
which provides a much better description of the data than its
counterpart for mesons.
The UCLA model and the Lund and Bowler functions also contain
$(1-z)^{a}$ terms.
For UCLA, the fitted value of $a\! =\! 2.9$ is close to 3, 
as anticipated~\cite{ucla}.

\begin{table}[t]
\caption{ \label{tab:modelresult} 
  Results of the fragmentation model tests.
  The minimum $\chi^{2}$ value, number of degrees of freedom, 
  fitted parameter values, 
  and the mean value of the corresponding scaled momentum distribution 
  are listed.
  }
\begin{center}
\begin{tabular}{l@{\hspace{0.5cm}}r@{\hspace{0.4cm}}l@{}c@{}ll@{\hspace{0.5cm}}c}
\hline\hline
                  &        &           &   &      &          &       \\[-0.3cm]
Model             & $\chi^{2}/dof$
                           & \multicolumn{4}{c}{Parameters} 
                           & $\left<x_p\right>$                      \\[0.05cm]
\hline
                  &        &           &   &      &          &       \\[-0.3cm]
JETSET$+$CS       & 227/39 & $\epsilon$&$=$& 0.135&          & 0.563 \\[0.05cm]
JETSET$+$BCFY     & 234/39 & $d$       &$=$& 0.355&          & 0.560 \\[0.05cm]
JETSET$+$KLP-M    & 219/39 & $\alpha$  &$=$& 3.05 &          & 0.572 \\[0.05cm]
JETSET$+$KLP-B    &  48/39 & $\alpha$  &$=$& 7.62 &          & 0.580 \\[0.05cm]
JETSET$+$Bowler   &  52/38 & $a$       &$=$& 0.93,& $b=0.88$ & 0.583 \\[0.05cm]
JETSET$+$Peterson & 100/39 & $\epsilon$&$=$& 0.077&          & 0.559 \\[0.05cm]
JETSET$+$Lund     &  49/38 & $a$       &$=$& 1.20,& $b=0.71$ & 0.584 \\[0.05cm]
UCLA              & 107/38 & $a$       &$=$& 2.9, & $b=0.74$ & 0.584 \\[0.05cm]
HERWIG            & 456/40 &           &---&      &          & 0.546 \\[0.05cm]
\hline\hline
\end{tabular}
\end{center}
\end{table}

The models predict $P(x_p)$ for primary leading charmed hadrons,
whereas the data also contain secondary charmed hadrons from the
splitting of hard gluons, 
and some of the reconstructed \Lc are decay products of other charmed baryons.
Both of these effects are included in the JETSET, HERWIG and UCLA
models, but it is important to consider the effects of possible
mismodelling.
The fraction of \qqbar events containing a gluon splitting
into a \ccbar pair has been measured at $\sqrt{s}\! =\!92~\gev$ to
be about 0.01~\cite{gtoccbar}.
At our lower c.m.\ energy this rate is expected to be reduced, and
the fraction of these that produce charmed baryons is expected to be
lower than that for primary $c$ and \cbar quarks.
The JETSET, UCLA and HERWIG models predict overall contributions of only
(0.009$\pm$0.004)\%, (0.017$\pm$0.005)\% and (0.034$\pm$0.012)\%, respectively,
concentrated at low $x_p$. 
The uncertainties in our lowest-$x_p$ bins are large enough to
accommodate such a contribution.
Adjusting the models to remove or double this contribution does not
change the results of the fits significantly.

Currently there are three known $\Sigma_c$ states 
(with masses 2455, 2520 and 2880~\mevcc) that decay into $\Lc \pi$, 
and four $\Lambda_c$ states (masses 2593, 2625, 2765 and 2880~\mevcc) 
that decay into $\Lc\pip\pim$.
In decays of such baryons into a slightly lighter baryon and one or two
pions, the daughter baryon carries most of the momentum, so the effect
of any such decay is to soften $P(x_p)$ slightly without distorting
it substantially.
In the JETSET$+$Lund, JETSET$+$Bowler and UCLA models, 
this effect is partially compensated by the fact that the 
heavier baryons are generated with slightly harder distributions.
Collectively, the effect is to broaden $P(x_p)$ slightly and shift
its average value down.
Combining our \Lc candidates with additional pions in the same event, 
we see clear signals for all of these states, and can compare the
relative contributions to the detected \Lc with the simulation.
The largest contribution of about 7\% from $\Sigma_c$(2455)
is well simulated in all models,
but the $\Sigma_c$(2520) rate is too high by a factor of $\sim$3, 
and the excited $\Lambda_c$ states are not in any simulation.
Removing two-thirds of the $\Sigma_c$(2520) narrows the simulated distributions
slightly, but does not improve any \chisq value significantly.
Similarly, adding excited $\Lambda_c$ states broadens all
distributions slightly, with no change in the conclusions of the model
tests.
No $\Xi_c$ or $\Omega_c$ states are known to decay to \Lc~\cite{pdg},
except two recently reported by Belle~\cite{bellexic}.
The latter are observed at very low rates, so should have negligible
effect on $P(x_p)$.


\section{\boldmath Results for $\Y4S$ Decays}
\label{sec:y4s}

\subsection{Charmed Baryon Production}

\begin{figure}[t]
\begin{center} 
\includegraphics[width=\linewidth]{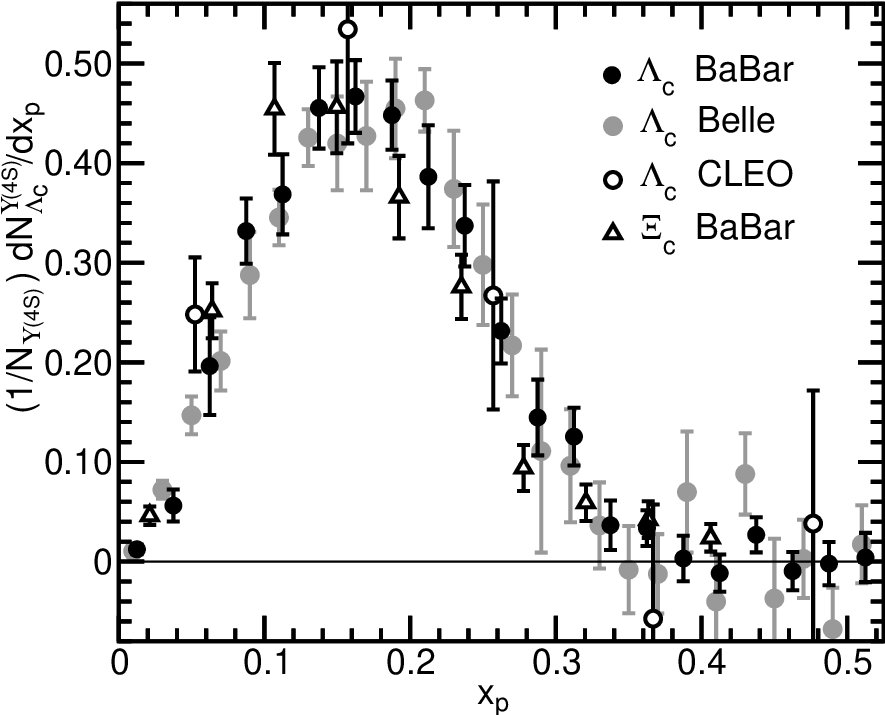}
\end{center}
\vspace{-0.7cm}
\caption {
  Differential \Lc production rate per \Y4S decay compared with
  previous measurements.
  Normalization errors are not shown, and 
  the \Xc0 rate is normalized to match the peak \Lc rate.
  } 
\label{fig:xsy4s}
\end{figure}

\noindent
The differential production rate per \Y4S decay is shown in 
Fig.~\ref{fig:xsy4s} and listed in Table~\ref{tab:xsy4s}.
The errors are as for the \qqbar results, with an additional
1.5\% normalization uncertainty due to the \qqbar subtraction 
procedure.
The kinematic limit for $\Y4S \!\!\to\! \BB \!\!\to\! \Lc\overline{p}$
decays is $x_p\! =\! 0.47$, and above this value the rate is
consistent with zero.
This region is omitted from the table and only partly shown in the figure.
We calculate the total rate by integrating the differential rate
over the kinematically allowed bins.
The resulting product of total rate and branching fraction,
$\NLcyy \cdot \BpKpi = 4.56 \pm 0.09\,\, {\rm (stat.)}
                           \pm 0.31\,\, {\rm (syst.)} \times 10^{-3}$,
is consistent with previous measurements~\cite{bellechsp,cleoBtocb}.
It corresponds to
$\NLcyy \! =\! 0.091 \pm 0.006$ (exp.)$\pm 0.024$ (BF) \Lc per \Y4S decay,
and 4.5$\pm$1.2\% of $B^0/B^-$ decays including a \Lc baryon,
assuming the \Y4S decays predominantly to \BB.

Our results on the shape are consistent with, and more precise than,
previous results, which are also shown in 
Fig.~\ref{fig:xsy4s}.
The $x_p$ distribution is quite soft.
In particular, the data drop rapidly above the peak and are 
consistent with zero above $x_p\!\! \approx\! 0.35$.
This is the range expected for quasi-two-body decays into a \Lc
or $\Sigma_c$ plus an antibaryon such as a $\overline{p}$, $\overline{n}$ or
$\overline{\Delta}$, and includes much of the range expected for
decays involving one or two additional pions.
The $\Bzb \!\!\to\! \Lc\overline{p}$ decay has been observed~\cite{bellelcp} 
at a very low rate consistent with our inclusive data.

A soft $x_p$ distribution was also seen in our recent study of \Xc0
baryons~\cite{bbrxic},
but the statistics of that study did not allow a meaningful direct
subtraction of the contribution from \qqbar events.
As an exercise, we assume a smooth distribution from \qqbar events
by choosing an empirical function that describes both the \Lc data in
Sec.~\ref{sec:shape} and the high-$x_p$ \Xc0 data (Fig.~\ref{fig:xsdlc}a).
We fit this function to the high-$x_p$ \Xc0 data and subtract the
result in all $x_p$ bins.
The resulting approximate $x_p$ distribution for \Xc0 in \Y4S decays
is also shown in Fig.~\ref{fig:xsy4s}, 
normalized to have roughly the same peak height as the \Lc data.
It is also quite soft, 
similar in shape to the \Lc but shifted slightly downward in $x_p$,
and consistent with zero above $x_p\!\! \approx\! 0.35$.
Because of the \eeccb subtraction procedure, the error bars cannot be
compared with those for the \Lc data, 
but the noted features do not depend on the details of the subtraction
procedure.

\begin{table}[t]
\caption{ \label{tab:xsy4s} 
  \Lc differential production rate per \Y4S decay per unit
  scaled momentum (to $\sqrt{s}\! =\! 10.58$~\gev).
  The last column includes those experimental errors that are
  correlated between $x_p$ values and affect the shape of the distribution.
  Normalization uncertainties are given only on the total.
  }
\begin{center}
\begin{tabular}{cr@{\hspace{0.4cm}}ccc}
\hline\hline
       &      &             &              &                \\[-0.3cm]
 $x_p$ &      & Statistical & \multicolumn{2}{c}{Systematic}\\[-0.35cm]
 Range & \multicolumn{1}{c}{\Large $\frac{1}{N_{\Upsilon}}
                                    \frac{d\NLcyy}{dx_p}$} &
 Error & Independent        & Correlated                  \\[0.25cm]
\hline
              &          &          &          &          \\[-0.3cm]
 0.000--0.025 &   0.0123 &   0.0023 &   0.0040 &   0.0013 \\
 0.025--0.050 &   0.0563 &   0.0073 &   0.0135 &   0.0046 \\
 0.050--0.075 &   0.1963 &   0.0110 &   0.0443 &   0.0182 \\
 0.075--0.100 &   0.3317 &   0.0152 &   0.0194 &   0.0215 \\
 0.100--0.125 &   0.3686 &   0.0193 &   0.0196 &   0.0294 \\
 0.125--0.150 &   0.4555 &   0.0208 &   0.0285 &   0.0205 \\
 0.150--0.175 &   0.4669 &   0.0221 &   0.0238 &   0.0167 \\
 0.175--0.200 &   0.4482 &   0.0201 &   0.0240 &   0.0151 \\
 0.200--0.225 &   0.3863 &   0.0211 &   0.0454 &   0.0132 \\
 0.225--0.250 &   0.3372 &   0.0204 &   0.0339 &   0.0109 \\
 0.250--0.275 &   0.2315 &   0.0197 &   0.0246 &   0.0084 \\
 0.275--0.300 &   0.1447 &   0.0183 &   0.0327 &   0.0064 \\
 0.300--0.325 &   0.1255 &   0.0171 &   0.0227 &   0.0055 \\
 0.325--0.350 &   0.0365 &   0.0175 &   0.0171 &   0.0041 \\
 0.350--0.375 &   0.0336 &   0.0160 &   0.0072 &   0.0037 \\
 0.375--0.400 &   0.0031 &   0.0163 &   0.0155 &   0.0043 \\
 0.400--0.425 &$-$0.0117 &   0.0163 &   0.0080 &   0.0042 \\
 0.425--0.450 &   0.0268 &   0.0150 &   0.0081 &   0.0044 \\
 0.450--0.475 &$-$0.0096 &   0.0158 &   0.0097 &   0.0048 \\
\hline
              &          &          &          &          \\[-0.3cm]
 Total        &   0.0910 &   0.0019 &   0.0026 &   0.0049 \\
 Norm. error  &   0.0029 &          &          &          \\
 BF error     &   0.0237 &          &          &          \\
\hline\hline
\end{tabular}
\end{center}
\end{table}

\subsection{Model Tests}

\begin{figure}[t]
\begin{center} 
\includegraphics[width=\linewidth]{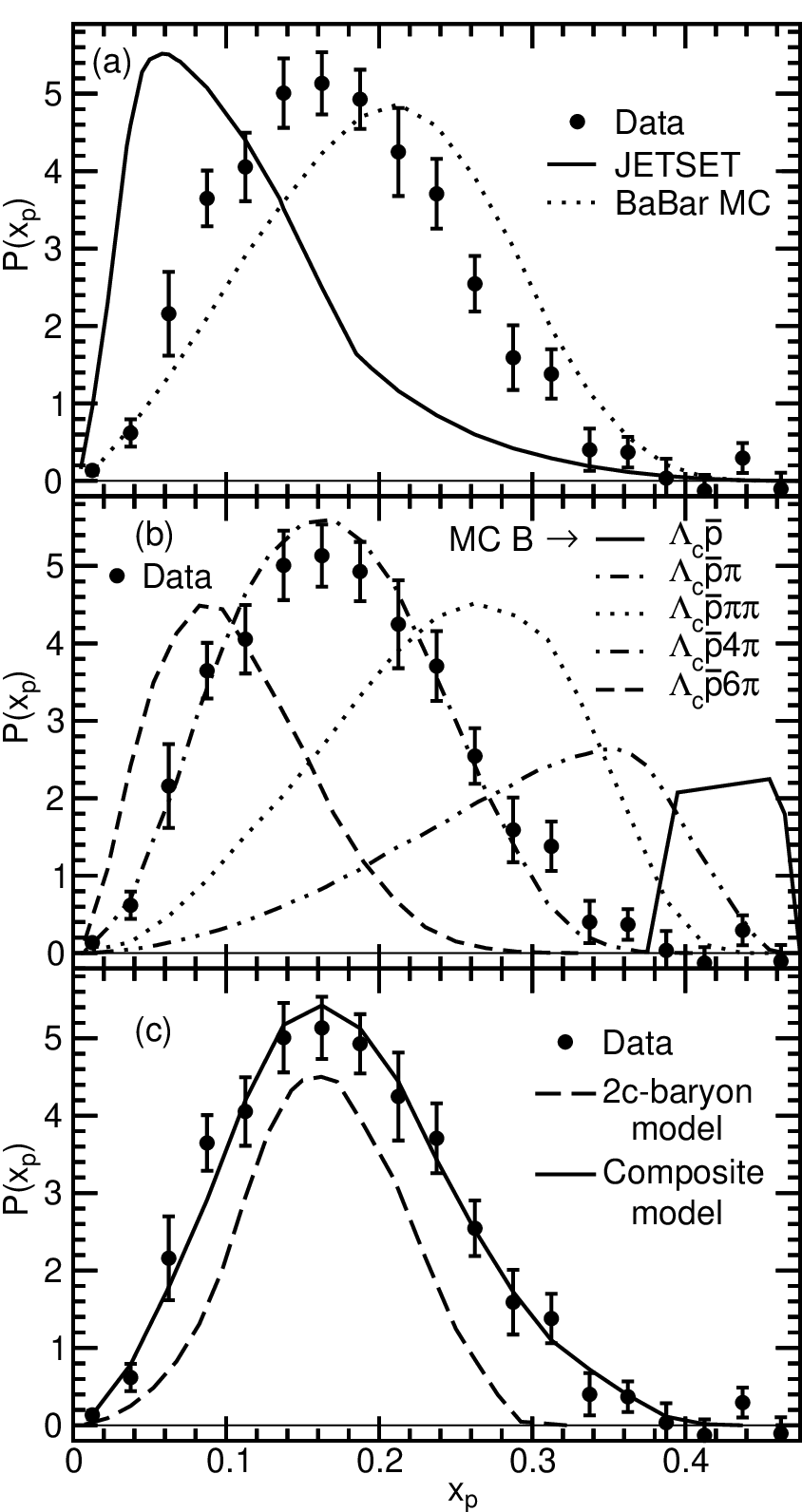}
\end{center}
\vspace{-0.5cm}
\caption {
  $P(x_p)$ for \Lc from \Y4S decays (dots) compared
  with the simulations (lines) described in the text.
  The data are normalized to unit area, as are the \babar\ simulation in (a),
  the $B \!\to \Lc\overline{p} 4\pi$ simulation in (b)
  and the composite simulation in (c);
  all other simulations are normalized arbitrarily.
  } 
\label{fig:xsdlcy}
\end{figure}

\noindent
Existing models of $B$ meson decays into $c$ baryons
were developed with little data on the $x_p$ distribution.
In Fig.~\ref{fig:xsdlcy}a we compare $P(x_p)$ (normalized over the
range $x_p\! <\! 0.475$) with the predictions of the models in the 
JETSET generator~\cite{jetset} and our internal generator~\cite{evtgen}.
The former has been tuned to measured multiplicities and
momentum distributions of stable $B$ decay products, 
and predicts a distribution that is much too soft.
The latter includes many measured exclusive decays involving $D$
mesons and postulates analogous few-body decays involving charmed baryons.
Its predicted $x_p$ distribution is similar in shape to the data, but 
shifted to higher $x_p$ values.
Neither generator produces simple (quasi-)two-body decays such as
$B \!\!\to\! \Lc\overline{p}$, $\Lc\overline{n}$, $\Lc\overline{\Delta}$
or $\Sigma_c\overline{p}$.

In Fig.~\ref{fig:xsdlcy}b we compare the same data with simulated
events of the type $B \!\!\to\! \Lc\overline{p}(m\pi)$ for selected
values of $m$, the number of pions in the decay in addition to the \Lc
and antiproton.
The distributions are insensitive to the charges of the pions or $B$
meson, or to replacing the antiproton with an antineutron.
Decays via a $\overline{\Delta}$ or strange antibaryon are not included;
they give \Lc distributions only slightly different from those shown with 
$m\! -\! 1$ pions.
For $m\! =\! 1$, 2, 4 and 6, the distributions shown are from the JETSET
simulation; 
phase space decays give similar distributions.
The spread in the distribution for $m\! =\! 0$ is due to the
finite momentum of the $B$ meson in the \epem c.m.\ frame.
The measured $P(x_p)$ is described adequately by the simulation with 
$m\! =\! 4$.
Adding contributions from $m\! =\! 3$ and $m\! =\! 5$ improves the 
\chisq of a comparison with the data, but no further contributions are helpful.
That $m$ is restricted to such a narrow range suggests that different
types of decay modes are needed.

An intriguing possibility is that there is a large contribution from
decays involving both a $c$ baryon and an anti-$c$ baryon.
The decays $\Bzb  \!\!\to\! \Xi_c^+ \Lcb$~\cite{bellexclc}
and $\Bm  \!\!\to\! \Lc \Lcb K^-$~\cite{bellelclck} have recently
been observed,
and we have previously measured an unexpectedly high rate of inclusive \Lcb in 
\Bm decays~\cite{bbrbtoccor}.
As an exercise, we model two-body two-$c$-baryon decays based on the simplest
internal $W$ diagrams, 
i.e.\ those of the forms
$\Bm  \!\!\to\! \Xi_c^+(\Lambda_c^+) \overline{\Sigma}_c^{--}$,
$\Bm  \!\!\to\! \Xi_c^0(\Sigma_c^0)  \overline{\Lambda}_c^{-}$,
$\Bzb \!\!\to\! \Xi_c^+(\Lambda_c^+) \overline{\Lambda}_c^{-}$, and
$\Bzb \!\!\to\! \Xi_c^0(\Sigma_c^0)  \overline{\Sigma}_c^{0}$,
where the parentheses indicate 
Cabibbo-suppressed modes.
Here, $\Xi_c$ represents any of the states $\Xi_c(2470)$,
$\Xi_c^{\prime}(2570)$ or $\Xi_c(2645)$,
$\Sigma_c$ represents $\Sigma_c(2455)$ or $\Sigma_c(2520)$,
$\Lambda_c$ represents $\Lambda_c(2285)$, $\Lambda_c(2593)$ or 
$\Lambda_c(2625)$, and
we consider all kinematically allowed combinations at relative rates determined
by phase space.
We decay all $\Sigma_c$ baryons into $\Lc \pi$ and all excited
$\Lambda_c$ baryons into $\Lc \pi\pi$, with 73\% of the $\Lambda_c^+(2593)$
baryons decaying through a $\Sigma_c\pi$ intermediate state and all
others via phase space.

The $x_p$ distribution of the \Lc from this simulation is compared with the 
data in Fig.~\ref{fig:xsdlcy}c.
Although it is too narrow to describe the data completely, it appears that
such processes could contribute substantially to the overall rate.
Combining this simulation with those for the 
$B \!\!\to\! \Lc\overline{p}(m\pi)$ modes
and assuming a smooth, broad distribution of $m$, 
we can describe the data with as much as a 50\% contribution from these
two-$c$-baryon decays.
For example, the ``composite model" in
Fig.~\ref{fig:xsdlcy}c comprises
35\% two-$c$-baryon decays
and (12, 25, 12, 9, 7)\% of $m\! =$(2, 3, 4, 5, 6),
and provides an excellent description of the data.
Decays with two charmed baryons and additional pions or kaons could
also contribute at low $x_p$, shifting the $m$ distribution downward.
Measurements of many exclusive baryonic $B$ decays, including both
one- and two-$c$-baryon modes, are needed to understand the
dynamics in detail.


\section{Summary and Conclusions}
\label{sec:summ}

\noindent
We use the excellent tracking and particle identification capabilities of
the \babar\ detector to reconstruct large, clean samples of \Lc baryons 
over the full kinematic range.
We measure their total production rates and inclusive scaled momentum
distributions 
in both \eehad events at $\sqrt{s}\! =\! 10.54$~\gev and \Y4S decays.
Our results are consistent with those published previously and
more precise.

In \eeqqb events we measure a total rate per event times branching fraction
into the \pKpi mode of
\begin{equation*}
\NLcqq \cdot \BpKpi = 2.84 \pm 0.04\,\, {\rm (stat.)}
                           \pm 0.09\,\, {\rm (syst.)} \times 10^{-3},
\end{equation*}
where the first error is statistical and the second systematic.
The uncertainty on the total rate per \qqbar event,
$\NLcqq \!\! =\! 0.057 \pm 0.002 \,{\rm (exp.)} \pm 0.015\!$ (BF),
is dominated by the uncertainty on \BpKpi.
The corresponding value of 
$N_{\Lambda c}^c\!\! = \! 0.071\pm$0.003 (exp.)$\pm$0.018 (BF)
\Lc per $c$-quark jet is
consistent with the hypothesis that roughly 10\% of $c$~jets produce a
$c$ baryon and a large fraction of these decay via a \Lc.

The scaled momentum distribution peaks at $x_p\!\approx\! 0.6$.
It is similar in shape to those measured previously for $D$ and $D^*$ mesons,
but peaks more sharply and drops toward zero more rapidly as $x_p \!\!\to\! 1$.
We measure an average value for \Lc of
\begin{equation*}
\left< x_p \right> = 0.574 \pm 0.006\,\, {\rm (stat.)}
                           \pm 0.006\,\, {\rm (syst.)},
\end{equation*}
which is consistent with values measured for ground state $D$
mesons, but about 5\% lower than those for $D^*$ mesons.
We use this distribution to test several models of heavy quark fragmentation,
none of which provides a complete description of the data.
The baryon-specific model of Kartvelishvili et al.\ and the models of
Lund and Bowler have acceptable \chisq values,
but all show a steeper slope on the low side of the peak than the data.
The UCLA model shows similar qualitative features, but worse agreement
with the data.
The HERWIG model is far too narrow, and all others are too broad.
In previous model tests using specific $c$ mesons~\cite{bellechsp}
and inclusive $b$ hadrons~\cite{bfrag} (a mix of roughly 90\% mesons 
and 10\% baryons),
the Lund, Bowler and Kartvelishvili models generally gave the best
description of the data, and UCLA described the $b$-hadron data,
whereas the other models 
showed discrepancies similar in form to those reported here.
The Kartvelishvili and UCLA models postulate different spectra for
mesons and baryons.
Their strong preference for their respective baryonic forms, combined
with the observed differences in shape between our \Lc spectrum and
previously measured $D$ meson spectra, indicate a difference in the 
underlying dynamics.

In \Y4S decays, we measure a total rate per event times branching fraction of 
\begin{equation*}
\NLcyy \cdot \BpKpi = 4.56 \pm 0.09\,\, {\rm (stat.)}
                           \pm 0.31\,\, {\rm (syst.)} \times 10^{-3},
\end{equation*}
corresponding to
$\NLcB \! =\! 0.045 \pm 0.003$ (exp.)$\pm 0.012$ (BF) per \Bzb/\Bub decay.
The spectrum is softer than predicted by our $B$ decay model, and
much harder than that predicted by JETSET.
It can be described by models of $B \!\!\to\! \Lc\overline{p}(m\pi)$
decays only if the $m\!=$ 3--5 contributions dominate.
Alternatively, a model including a large contribution from decays
involving both a charmed and an anti-charmed baryon can describe the
data in conjunction with a broad distribution of $m$.
Additional studies of exclusive modes are needed to understand the
details of $B$-meson decays into baryons.

\vspace*{0.50truecm}
\section*{Acknowledgements}

\noindent
We are grateful for the 
extraordinary contributions of our \pep2\ colleagues in
achieving the excellent luminosity and machine conditions
that have made this work possible.
The success of this project also relies critically on the 
expertise and dedication of the computing organizations that 
support \babar.
The collaborating institutions wish to thank 
SLAC for its support and the kind hospitality extended to them. 
This work is supported by the
US Department of Energy
and National Science Foundation, the
Natural Sciences and Engineering Research Council (Canada),
Institute of High Energy Physics (China), the
Commissariat \`a l'Energie Atomique and
Institut National de Physique Nucl\'eaire et de Physique des Particules
(France), the
Bundesministerium f\"ur Bildung und Forschung and
Deutsche Forschungsgemeinschaft
(Germany), the
Istituto Nazionale di Fisica Nucleare (Italy),
the Foundation for Fundamental Research on Matter (The Netherlands),
the Research Council of Norway, the
Ministry of Science and Technology of the Russian Federation, 
Ministerio de Educaci\'on y Ciencia (Spain), and the
Particle Physics and Astronomy Research Council (United Kingdom). 
Individuals have received support from 
the Marie-Curie IEF program (European Union) and
the A. P. Sloan Foundation.



\end{document}